\documentclass[twocolumn,showpacs,prb,aps]{revtex4}
\usepackage{amsmath}
\usepackage{amssymb}
\usepackage{graphicx}
\usepackage{dcolumn}
\usepackage{bm}
\usepackage[dvips]{color}

\begin{document}

\title{Anomalous Enhancement of the Superconducting Transition Temperature
in Electron-Doped Cuprate Heterostructures}
\author{ K. Jin$^{1}$, P. Bach$^{1}$, X. H. Zhang$^{1}$, U. Grupel$^{2}$, E.
Zohar$^{2}$, I. Diamant$^{2}$, Y. Dagan$^{2}$, S. Smadici$^{3}$, P. Abbamonte%
$^{3}$, and R. L. Greene$^{1}$}
\affiliation{$^{1}$Center for Nanophysics $\And $ Advanced Materials and Department of
Physics, University of Maryland, College Park, MD 20742, USA\\
$^{2}$Raymond and Beverly Sackler School of Physics and Astronomy, Tel-Aviv
University, Tel Aviv 69978, Israel\\
$^{3}$Frederick Seitz Materials Research Laboratory, University of Illinois,
Urbana, IL 61801, USA}
\date{\today }

\begin{abstract}
The superconducting transition temperature $T_{c}$ of multilayers of
electron-doped cuprates, composed of underdoped (or undoped) and overdoped La%
$_{2-x}$Ce$_{x}$CuO$_{4}$ (LCCO) and Pr$_{2-x}$Ce$_{x}$CuO$_{4}$ (PCCO) thin
films, is found to increase significantly with respect to the $T_{c}$ of the
corresponding single-phase films. By investigating the critical current
density of superlattices with different doping levels and layer thicknesses,
we find that the $T_{c}$ enhancement is caused by a redistribution of charge
over an anomalously large distance.
\end{abstract}

\pacs{74.78.Fk, 74.72.Ek, 74.72.-h, 74.25.Sv}
\maketitle

One of the striking properties of the high--$T_{c}$ cuprates is the strong
dependence on the number of charge carriers put into the CuO$_{2}$ planes of
many of their electronic properties. For example, for low carrier
concentration (underdoped regime) the material is insulating; it is metallic
for high carrier concentration (overdoped regime), and superconducting in
between, with a maximal $T_{c}$ for the optimum doping. The charge carriers
can be either holes or electrons.\cite{GreeneRMP}

Recently, it has been shown that heterostructure bi-layers (BL's) and
super-lattices (SL's), composed of metallic and insulating hole-doped
cuprates, can be superconducting.\cite{Bozovic2008,Smadici2009} Also, a BL
of superconducting La$_{2-x}$Sr$_{x}$CuO$_{4}$ (LSCO) and metallic,
nonsuperconducting La$_{1.65}$Sr$_{0.35}$CuO$_{4}$ has a critical
temperature ($T_{c}$) greater than the superconducting single-phase LSCO
itself, with a maximal enhancement for $x=0.125$.\cite{Yuli2008,Millo2010}
In this case, the superconducting layer was shown to be confined to the
interface. The $T_{c}$ enhancement in Refs.4 and 5 has been interpreted in
terms of an interplay between the large pairing amplitude in the underdoped
(UD) cuprate and the phase stiffness originating from the metallic overdoped
(OD) layer.\cite{Kivelson2008} In this model, $T_{c}$ in UD cuprates is not
the temperature where the pairs are formed and condensed as in the BCS
theory, but a temperature at which phase order is destroyed by fluctuations
despite the persistence of pairing amplitude.\cite{Kivelson1995} Other
possible explanations for the $T_{c}$ enhancement are: oxygen diffusion or
uptake, strain induced by the substrate \cite{Locquet1998} and cationic
diffusion between the layers. The latter was shown to be negligible for LSCO.%
\cite{Bozovic2008}

In hole-doped cuprates, a large phase fluctuation region above $T_{c}$ has
been observed by different techniques such as the Nernst effect and torque
magnetization,\cite{Wang2006,Li2010} and scanning tunneling microscopy
measurements.\cite{Gomes2007,Yuli2009} In contrast, in the electron doped
cuprates, the pairing amplitude seems to follow $T_{c}$ even for the UD side,%
\cite{Dagan2009} and phase fluctuations appear to be small as indicated by
the narrow region of vortex Nernst effect observed above $T_{c}$.\cite%
{Li2007} The normal state tunneling gap was interpreted as preformed
superconductivity, however, this gap vanished at rather low temperature, not
very far away from $T_{c}$ for UD samples.\cite{Dagan2005} If the interplay
between the pairing amplitude in the UD layer and the phase stiffness in the
OD layer is the origin of the $T_{c}$ enhancement in cuprate SL's one would
expect a smaller enhancement for electron-doped heterostructures.

Here we report a large $T_{c}$ enhancement in electron-doped cuprate
heterostructures of both La$_{2-x}$Ce$_{x}$CuO$_{4}$ (LCCO) and Pr$_{2-x}$Ce$%
_{x}$CuO$_{4}$ (PCCO) SL's. This large $T_{c}$ enhancement argues against an
increased phase stiffness in the UD layer as proposed in Ref.6. However, we
find that the critical current of the SL's scales with the total layer
thickness and not with the number of interfaces. This suggests that the
observed $T_{c}$ enhancement is not confined to the interfaces but is caused
by a global redistribution of charge carriers. In addition, we find that the
$T_{c}$ of the SL's scales with the $c$-axis lattice parameter as found in
single-phase PCCO films.\cite{Maiser1998} From these results we conclude
that the $T_{c}$ enhancement in electron-doped SL's is caused by an
anomalous charge redistribution over a very large length scale of at least
20 nm. Our results suggest that the $T_{c}$ enhancement found in hole-doped
cuprates may have the same origin.

%
\begin{figure*}[tbp]
\begin{center}
\includegraphics*[bb=99 506 514 717, width=10cm,clip]{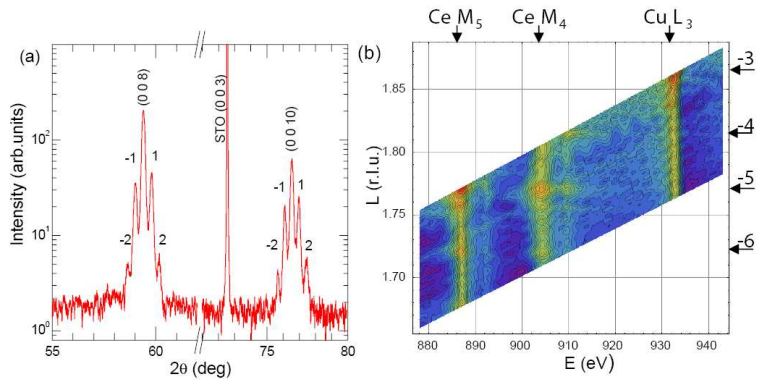}
\end{center}
\caption{(color online) (a) XRD $\protect\theta -2\protect\theta $ spectra
of (0.19/0.06)$_{n=10}$. The main Bragg peaks of (0 0 8) and (0 0 10) and
the superlattice satellite peaks are marked. (b) Energy dependence of SL
reflections near Ce and Cu edges measured with resonant soft x-ray
scattering for sample (0.19/0.06)$_{n=10}$. Horizontal arrows (with -3, -4,
-5, -6) point to positions of superlattice peaks predicted from XRD
measurements. Peaks are clearly visible at Ce M$_{4}$ and M$_{5}$ edges
around 910 eV and 890 eV respectively, but not at the Cu L$_{3}$ edge.
Shorter period oscillations in the figure are fringes due to finite sample
thickness. The soft x-ray energy limits the scattering momentum at high L.}
\end{figure*}
%

Epitaxial films were grown on SrTiO$_{3}$ (STO) substrates by the pulsed
laser deposition (PLD) method.\cite{jk2009,Dagan2004} We took extra care
optimizing the annealing conditions for each geometry and composition to
obtain the maximal $T_{c}$. LCCO is chemically stable for $x\geq $0.06
whereas it has a contaminative T-phase for lower Ce concentrations.\cite%
{Naito2002,Sawa2002} PCCO can be deposited with $x<0.20$. SL's of LCCO were
prepared with successive layers of $x=0.06$ (UD) and OD layer with $%
x=0.17,0.19,$ and 0.21. For PCCO SL's we used consecutive layers of $x=0.19$
(OD)\ and PCCO with $x=0.00,0.11$ and 0.12. BL's of PCCO are composed of $%
x=0.12$ and $x=0.19$. For all the SL's in this study the total thickness is
kept $\sim 140$ nm, and each layer has the same thickness. We use the
notation ($x$/0.06)$_{n}$ for $n$ interfaces between alternate OD LCCO ($%
x=0.17,0.19,$ and 0.21$)$ and UD $x=0.06$, as seen in the inset of Fig. 3(c)
for (0.19/0.06)$_{n}$.

Figure 1(a) presents x-ray $\theta -2\theta $ diffraction (XRD) spectra of
LCCO sample (0.19/0.06)$_{n=10}$, which has 11 layers with each layer $\sim
13$ nm ($\sim $ 10 u.c.). Well-defined multiple satellite peaks originated
from the modulated structure can be observed along with the main peaks of
single-phase films. This confirms the periodic layered structure. The
modulation period can be calculated from $\Lambda =\lambda /2(\sin \theta
_{i}-\sin \theta _{i+1})$, where $i$ and $i+1$ represent two nearest
satellite peaks.\cite{Zhu2007} From our data we obtain $\Lambda \sim 26\ $%
nm, consistent with the expected 20 u.c. periodicity from the deposition
rate. Resonant soft x-ray scattering \cite{Smadici2009} for (0.19/0.06)$%
_{n=10}$ sample show a SL modulation at the Ce edges. As seen in Fig. 1(b),
several satellite peaks that coincide with the positions predicted from the
above equation with $i=-3,-4,-5,-6$ are clearly visible at $\sim $ 910 eV
(Ce M$_{4}$ edge) and 890 eV (Ce M$_{5}$ edge), respectively. This
observation indicates that the SL reflections really arise from a selective
modulation in the Ce density, whereas the Cu atoms reside in their usual,
lattice positions. We also did an electron energy-loss spectroscopy (EELS)
measurement on a similar SL film and found a clear Ce modulation.

Figure 2(a) shows one example of $T_{c}$ enhancement in a LCCO SL,
(0.19/0.06)$_{n=10}$. The $x=0.06$ single-phase films have a broad
transition with onset transition temperature $T_{c}^{onset}<12$ K; the OD $%
x=0.19$ is metallic and nonsuperconducting. After constructing the SL, the $%
T_{c}^{onset}$ of (0.19/0.06)$_{n=10}$ reaches up to 24 K, comparable to the
value of optimal doping $x=0.11$ with $T_{c}^{onset}\sim 26$ K. The real
part of the ac susceptibility for (0.19/0.06)$_{n=10}$ and single-phase $%
x=0.11$ are shown in the inset of Fig. 2(a). In Fig. 2(b) and (c), the
transition temperatures $T_{c0}$\ (zero resistance of superconducting
transition) and $T_{c}^{onset}$ are plotted against the doping for LCCO and
PCCO, respectively. All the SL's show an obvious $T_{c}$ enhancement with
respect to the single-phase films themselves$-$almost constant for either
LCCO with varying OD layer or PCCO with varying UD layer$-$and are higher
than that of BL's. Since for the SL the $T_{c}$ enhancement is not small,
the question naturally arises as to what mechanism is responsible for this
enhancement.

One possible explanation is a substrate strain effect, which is expected to
be important in very thin films.\cite{Locquet1998,Bozovic2002} In our
experiment the thickness of the bottom layer changed from 5 nm to 20 nm
while $T_{c}$ remained approximately constant. It is therefore unlikely that
the enhancement is caused by strain.

Next, we check if the $T_{c}$ enhancement in the SL is confined to the
interface. Assuming well defined superconducting interfaces and a
non-superconducting bulk film, the super-current should flow along the
interfaces and \textquotedblleft short-circuit\textquotedblright\ the entire
film. In this scenario, one expects the critical current (I$_{c})$ to scale
with $n$, the number of the interfaces.\cite{Ref} In Fig. 3(a), typical
current (I) - voltage (V) characteristics are shown for LCCO (0.19/0.06)$%
_{n=40}$ at various temperatures. J$_{c}(T)$ is calculated from I$%
_{c}/(w\times t$) with $w$ the width of the current bridge and $t$ the total
thickness of the film. Since both the $w$ and $t$ are kept the same for
different SL's, then J$_{c}(T)$ should be proportional to $n$ in the
interface-effect scenario. We repeat these measurements for various $n$'s.
In particular we measured LCCO (0.19/0.06)$_{n}$ with $n=40,20,10,6$. These
SL's have approximately the same $T_{c}$. Surprisingly, at all temperatures J%
$_{c}$ is approximately independent of $n$ as seen in Fig. 3(b). In Fig.
3(c) we plot J$_{c}$ for various $n$'s against the reduced temperature ($%
1-T/T_{c0}$). It can be clearly seen that J$_{c}$\ is not proportional to $n$
but approximately the same, so the $T_{c}$ enhancement is not confined to
the interface! This can also rule out that the $T_{c}$ enhancement is caused
by slight Ce diffusion, i.e., several unit cells at the interface. Note that
our x-ray data [Fig. 1(b)] and EELS data (not shown) have excluded the case
of heavy Ce diffusion since Ce modulation is clearly observed.

%
\begin{figure}[tbp]
\begin{center}
\includegraphics*[bb=25 251 596 720, width=6.5cm,clip]{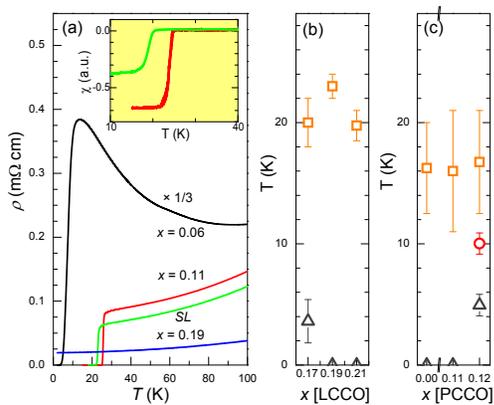}
\end{center}
\caption{(color online) (a) The resistivity, $\protect\rho (T)$, of
single-phase LCCO with $x=0.06,0.11,$ and 0.19, and also the (0.19/0.06)$%
_{n=10}$ SL. Note that $\protect\rho (T)$ of $x=0.06$ is divided by a factor
of 3.\ (Inset) The real part of ac susceptibility vs temperature for
single-phase LCCO with $x=0.11$ (red) and (0.19/0.06)$_{n=10}$ (green). (b) $%
T_{c}$ of single-phase LCCO ($\triangle $), ($x$/0.06)$_{n=10}$ SL's ($%
\square $). (c) $T_{c}$ of single-phase PCCO ($\triangle $), SL's with fixed
OD $x=0.19$ ($\square $) and BL's ($\bigcirc $). The BL data is obtained
from the structure of 240 nm $x=0.12$ and 60 nm $x=0.19$. $T_{c}$ of
single-phase PCCO with $x=0.19$ is $\sim $ 7.3 K. The top and bottom of the
error bars correspond to $T_{c}^{onset}$ and $T_{c0}$, respectively. }
\end{figure}
%

It has been reported that in La$_{2}$CuO$_{4}$ (LCO)/LSCO\ SL the conducting
holes do not follow the distribution of the Sr atoms, and the LCO layers
become highly doped. This suggests a redistribution of the charge carriers
among the layers.\cite{Smadici2009} A similar effect must take place in our
films, where, the charge carriers redistribute between the OD and UD layers.
In the charge-redistribution scenario, the total effective thickness of the $%
T_{c}$-enhanced region is determined by the Ce difference in the adjacent
layers, irrespective of $n$. Exactly how the charge redistributes is not
known but our data in Fig. 3(c) suggest that this charge is spread over at
least 20 nm since the J$_{c}$ of (0.19/0.06)$_{n}$ is approximately the same
for all $n$'s ( for $n=6$, the layer thickness is $20$ nm). Fig. 3(d) shows
the J$_{c}$ data for SL's $(x/0.06)_{n=10}$ with $x=0.17,0.19,$ and 0.21,
and also the single-phase $x=0.11$ film. We note that as $x$ increases the J$%
_{c}$ of $(x/0.06)_{n=10}$ decreases. This is understandable in the
charge-redistribution scenario since the $T_{c}$-enhanced region with
optimal charge count should be smaller as the Ce range becomes larger for
the same total film thickness. Future $T_{c}$ and J$_{c}$ experiments will
be needed on thicker layers to determine a more definitive number for the
anomalous length scale. We have attempted to make superlattices with $n=4$
and $2$, but so far without success.

%
\begin{figure}[tbp]
\begin{center}
\includegraphics[bb=55 367 395 695, width=6cm,clip]{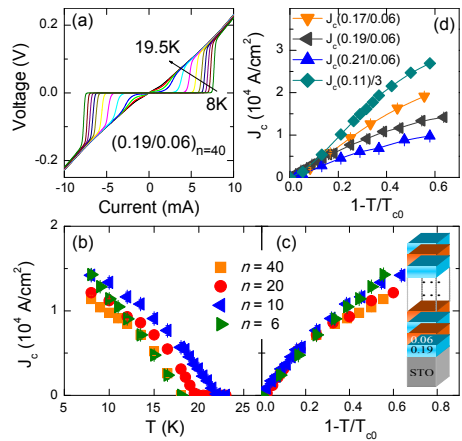}
\end{center}
\caption{(color online) (a) I-V of LCCO SL (0.19/0.06)$_{n=40}$ from 8 to
19.5 K. Critical currents were determined using a 1$\times $10$^{-6}$ V
criteria. (b) and (c) The critical current density J$_{c}$ versus $T$ and (1-%
$T/T_{c0})$ of LCCO SL's (0.19/0.06)$_{n}$ with $n=40,20,10,6,$
respectively. (Inset) The structure configurations for (0.19/0.06)$_{n}$.
(d) J$_{c}$ versus (1-$T/T_{c0})$ of LCCO single-phase $x=0.11$ (divided by
a factor of 3) and SL's ($x$/0.06)$_{n=10}$ with $x=0.17,0.19,$ and 0.21. }
\end{figure}
%

Fig. 4 presents $T_{c}$ versus \textit{c} axis lattice parameter $(c_{0})$
for LCCO films deposited in various techniques and SL's with different
annealing time. $T_{c}$ decreases on the overdoped side as $c_{0}$
decreases, both for SL's and single phase films. Similar results were shown
for single phase PCCO films with various Ce concentrations.\cite{Maiser1998}
Since Ce$^{4+}$ ion is smaller than Pr$^{3+}$, this can explain the decrease
of $c_{0}$ with increasing Ce doping as observed by Maiser \textit{et al.}.%
\cite{Maiser1998} In our case we observe a systematic change of $c_{0}$ with
$T_{c}$. This can be explained by charge redistribution in the SL's causing
an effect similar to that of Ce substitution in the single phase films.
Since XRD is a bulk measurement the systematic variation of $c_{0}$ gives
further indication that the charge redistribution occurs in the entire film
thickness.

Oxygen diffusion is one plausible explanation for this charge redistribution
over a very large distance. Unfortunately, the oxygen content in a single
phase thin film or in various parts of a SL or BL cannot be precisely
measured. Oxygen diffusion from UD to OD layers would increase $T_{c}$ since
this would add carriers (electrons) to the UD layer. Higgins et al.\cite%
{Higgins2006} demonstrated that such a process increased the carrier
concentration in the UD layers and decreased it in the OD ones. In that
case, the $T_{c}$ enhancement will take place only in the UD layers, because
for the overdoped layers the effect of carrier reduction is overwhelmed by
the disorder introduced by the apical oxygen (the site of the added oxygen)
resulting in a total decrease of $T_{c}$.\cite{Higgins2006} So, oxygen
diffusion from UD to OD layers appears to be a plausible explanation for the
global charge redistribution. However, the oxygen diffusion length is of the
order of the film thickness for the annealing temperature and time used in
our deposition. Thus, it is difficult to explain using an oxygen diffusion
scenario the difference between the BL, where the $T_{c}$ enhancement is
relatively small, and the SL $T_{c}$ enhancement. Therefore, it is possible
that the charge redistribution may have a different origin.

%
\begin{figure}[tbp]
\begin{center}
\includegraphics*[bb=94 306 453 656, width=4.5cm,clip]{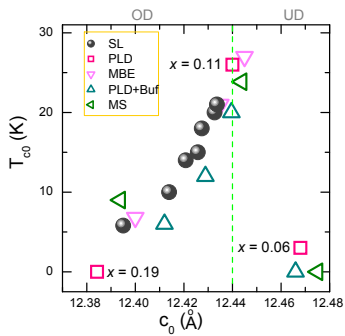}
\end{center}
\caption{(color online) The $c$-axis lattice parameter dependence of $T_{c0}$
for LCCO single-phase films prepared by PLD ($\square $), Molecular beam
epitaxy ($\bigtriangledown $) \protect\cite{Naito2002}, PLD with buffer
layer BaTiO$_{3}$($\triangle $) \protect\cite{Sawa2002}, and magnetron
sputtering ($\lhd $) \protect\cite{WBX2009}, and also for (0.19/0.06)$_{n}$
SL's with different $T_{c}$ ($\bigcirc $). The dashed line marks the optimal
doping.UD and OD represent underdoped and overdoped single-phase LCCO,
respectively.}
\end{figure}
%

We tried to estimate the charge redistribution length scale ($d$) in a SL
using a naive calculation. If we allow charge to move from the OD layer to
the UD one, the electrostatic energy increases. However, the free energy can
still decrease due to the enhanced superconductivity. Minimizing the free
energy yields a length scale of the order of $d\approx \frac{H_{c}}{16\pi
\rho }\sim $ 0.2 \AA , using the thermodynamic critical field H$_{c}=0.5$ T
and the charge density $\rho \approx 0.04$ electrons$/$Cu (or holes$/$Cu).
This means that the electrostatic energy is too high to explain the large
length scale over which the charge redistributes in our experiment.

Another possibility is that charge redistribution occurs on a length scale
determined by the screening length ($\lambda _{TF}$). Using the Thomas-Fermi
approximation we obtains a length scale of $\sim 10$ \AA , which is two
orders of magnitude smaller than what we observe. Certainly, this
approximation is oversimplified for the cuprates. Yet, in the hole-doped
LCO/LSCO SL an analysis of RSXS data yielded a screening length of $\sim 6$
\AA ,\cite{Smadici2009} not very different from our electron-doped results.
So both the naively calculated $\lambda _{TF}$ and $d$ length scales are
small and cannot explain the anomalously large charge redistribution length
that is inferred by our experiments.

In conclusion, we fabricated electron-doped superlattice and bilayer films
comprised of underdoped (or undoped) /overdoped La$_{2-x}$Ce$_{x}$CuO$_{4}$
(LCCO) and Pr$_{2-x}$Ce$_{x}$CuO$_{4}$ (PCCO), and find an enhanced $T_{c}$
which is comparable to optimally doped single-phase films for both LCCO and
PCCO. Our results suggest that a charge carrier redistribution occurs on an
anomalously large length scale of $\gtrsim $20 nm. This charge
redistribution is the cause for the $T_{c}$ enhancement. The exact mechanism
of the charge redistribution is not known, although oxygen diffusion is the
most plausible explanation. Further theoretical and experimental studies
will be needed to understand these remarkable results.

The work at UMD is supported by NSF Grant DMR-0653535 and AFOSR-MURI Grant
FA9550-09-1-0603, the work at Tel-Aviv and UMD by BSF Grant 2006385. RSXS
studies were supported by the U.S. Department of Energy Grant No.
DE-FG02-06ER46285, with use of the NSLS supported under Contract No.
DE-AC02-98CH10886.



\begin{thebibliography}{99}
\bibitem{Ref} I$_{c}=$J$_{c}^{^{\prime }}\times t^{^{\prime }}\times w\times
n$. Here J$_{c}^{^{\prime }}$ is the critical current density of the $T_{c}$%
-enhanced part, which has effective thickness $t^{^{\prime }}.$In the
interface-effect scenario, $t^{^{\prime }}$should be determined by the
proximity effect between adjacent layers, which is constant if the
constituents are fixed. Thus, I$_{c}\varpropto n$.
\end{thebibliography}
\end{document}